\documentclass[manuscript=article,journal=jpcl]{achemso}
\usepackage[dvipsnames]{xcolor}
\usepackage{subcaption}
\usepackage{amssymb}
\usepackage{amsmath}
\usepackage{braket}
\usepackage{multirow}
\usepackage{array}
\usepackage{adjustbox}
\usepackage{tablefootnote}
\newcolumntype{L}[1]{>{\raggedright\let\newline\\\arraybackslash\hspace{0pt}}m{#1}}
\newcolumntype{C}[1]{>{\centering\let\newline\\\arraybackslash\hspace{0pt}}m{#1}}
\newcolumntype{R}[1]{>{\raggedleft\let\newline\\\arraybackslash\hspace{0pt}}m{#1}}
\usepackage[version=3]{mhchem}
\usepackage{xr}

\makeatletter
\newcommand*{\addFileDependency}[1]{
  \typeout{(#1)}
  \@addtofilelist{#1}
  \IfFileExists{#1}{}{\typeout{No file #1.}}
}
\makeatother

\newcommand*{\myexternaldocument}[1]{%
    \externaldocument{#1}%
    \addFileDependency{#1.tex}%
    \addFileDependency{#1.aux}%
}

\myexternaldocument{SI}

\title{AtomSets - A Hierarchical Transfer Learning Framework for Small and Large Materials Datasets}
\author{Chi Chen}
\affiliation[UCSD]{Materials Virtual Lab, Department of NanoEngineering, University of California San Diego, 9500 Gilman Dr, Mail Code 0448, La Jolla, CA 92093-0448, United States}

\author{Shyue Ping Ong}
\affiliation[UCSD]{Materials Virtual Lab, Department of NanoEngineering, University of California San Diego, 9500 Gilman Dr, Mail Code 0448, La Jolla, CA 92093-0448, United States}
\email{ongsp@eng.ucsd.edu}

\date{}

\begin{document}

\maketitle
\newpage
\begin{abstract}
Predicting materials properties from composition or structure is of great interest to the materials science community. Deep learning has recently garnered considerable interest in materials predictive tasks with low model errors when dealing with large materials data. However, deep learning models suffer in the small data regime that is common in materials science. Here we leverage the transfer learning concept and the graph network deep learning framework and develop the AtomSets machine learning framework for consistent high model accuracy at both small and large materials data. The AtomSets models can work with both compositional and structural materials data. By combining with transfer learned features from graph networks, they can achieve state-of-the-art accuracy from using small compositional data ($<$400) to large structural data ($>$130,000). The AtomSets models show much lower errors than the state-of-the-art graph network models at small data limits and the classical machine learning models at large data limits. They also transfer better in simulated materials discovery process where the targeted materials have property values out of the training data limits. The models require minimal domain knowledge inputs and are free from feature engineering. The presented AtomSets model framework opens new routes for machine learning-assisted materials design and discovery. 

\end{abstract}

\section{Introduction}
Machine learning (ML) has garnered substantial interest as an effective method for developing surrogate models for materials property predictions in recent years.\cite{butlerMachineLearningMolecular2018,chenCriticalReviewMachine2020} However, a critical bottleneck is that materials datasets are often small and inhomogeneous, making it challenging to train reliable models. While large density functional theory (DFT) databases such as the Materials Project,\cite{jainCommentaryMaterialsProject2013} Open Quantum Materials Database\cite{kirklinOpenQuantumMaterials2015} and AFLOWLIB\cite{curtaroloAFLOWLIBORGDistributed2012} have $\sim$O($10^6$) relaxed structures and computed energies, data on other computed properties such as band gaps, elastic constants, dielectric constants, etc. tend to be several times or even orders of magnitude fewer.\cite{chenCriticalReviewMachine2020} In general, deep learning models based on neural networks tend to require much more data to train, resulting in lower performance in small datasets relative to non-deep learning models. For example, \citet{dunnBenchmarkingMaterialsProperty2020} have found that while deep learning models such as the MatErials Graph Networks (MEGNet)\cite{chenGraphNetworksUniversal2019} and Crystal Graph Convolutional Neural Network (CGCNN)\cite{xieCrystalGraphConvolutional2018} achieve state-of-the-art performance for datasets with $>$ O($10^4$) data points, ensembles of non-deep-learning models (using AutoMatminer) outperform these deep learning models when the data set size is $<$ O($10^4$), and especially when the data set is $<$ O($10^3$).

Several approaches have been explored to address the data bottleneck. The most popular approach is transfer learning (TL), wherein the weights from models trained on a property with a large data size are ``transferred'' to a model on smaller data size. Most TL studies were performed on the same property. For example, \citet{hutchinsonOvercomingDataScarcity2017} developed three TL approaches that reduced the model errors in predicting experimental band gaps by including DFT band gaps. Similarly, \citet{jhaEnhancingMaterialsProperty2019} trained models on the formation energies in the large OQMD database and demonstrated that transferring the model weights from OQMD can improve the models on the small DFT-computed and even experimental formation energy data. TL has also been demonstrated between different properties in some cases. For example, the present authors\cite{chenGraphNetworksUniversal2019} found that transferring the weights from large data-size formation energy MEGNet models to smaller-data-size band gap and elastic moduli models improved convergence rate and accuracy. Another approach uses multi-fidelity models, where datasets of multiple fidelities (e.g., band gaps computed with different functionals or measured experimentally) are used to improve prediction performance on the more valuable, high fidelity properties. For example, two-fidelity co-kriging methods have demonstrated successes in improving the predictions of the Heyd-Scuseria-Ernzerhof (HSE)\cite{heydHybridFunctionalsBased2003} band gaps of perovskites\cite{pilaniaMultifidelityMachineLearning2017}, defect energies in hafnia\cite{batraMultifidelityInformationFusion2019} and DFT bulk moduli\cite{batraMachineLearningMultifidelity2020}. In a recently published work, the present authors also developed multi-fidelity MEGNet models that utilize band gap data from four DFT functionals (Perdew-Burke-Ernzerhof\cite{perdewGeneralizedGradientApproximation1996} or PBE, Gritsenko-Leeuwen-Lenthe-Baerends with solid correction\cite{gritsenkoSelfconsistentApproximationKohnSham1995,kuismaKohnShamPotentialDiscontinuity2010} or GLLB-SC, strongly constrained and appropriately normed\cite{sunStronglyConstrainedAppropriately2015} or SCAN and HSE\cite{heydHybridFunctionalsBased2003}) and experimental measurements to significant improve the prediction of experimental band gaps.\cite{chenLearningPropertiesOrdered2021}

In this work, we develop ``AtomSets'', a hierarchical framework to TL using MEGNet models that can achieve uniformly excellent performance across diverse datasets with different sizes. The AtomSets framework unifies compositional and structural features under one umbrella. We show, for the first time, TL from structural models to compositional models. Using 13 MatBench datasets\cite{dunnBenchmarkingMaterialsProperty2020}, we show that the AtomSets models can achieve excellent performance even when the inputs are compositional and the data size is small ($\sim$ 300), while retaining MEGNet's state-of-the-art performance on properties with large data sizes. Furthermore, the model construction requires minimal domain knowledge and no feature engineering. 

\section{Methods}

\subsection{MatErials Graph Network}

\begin{figure}
\centering
\includegraphics[width=0.9\textwidth]{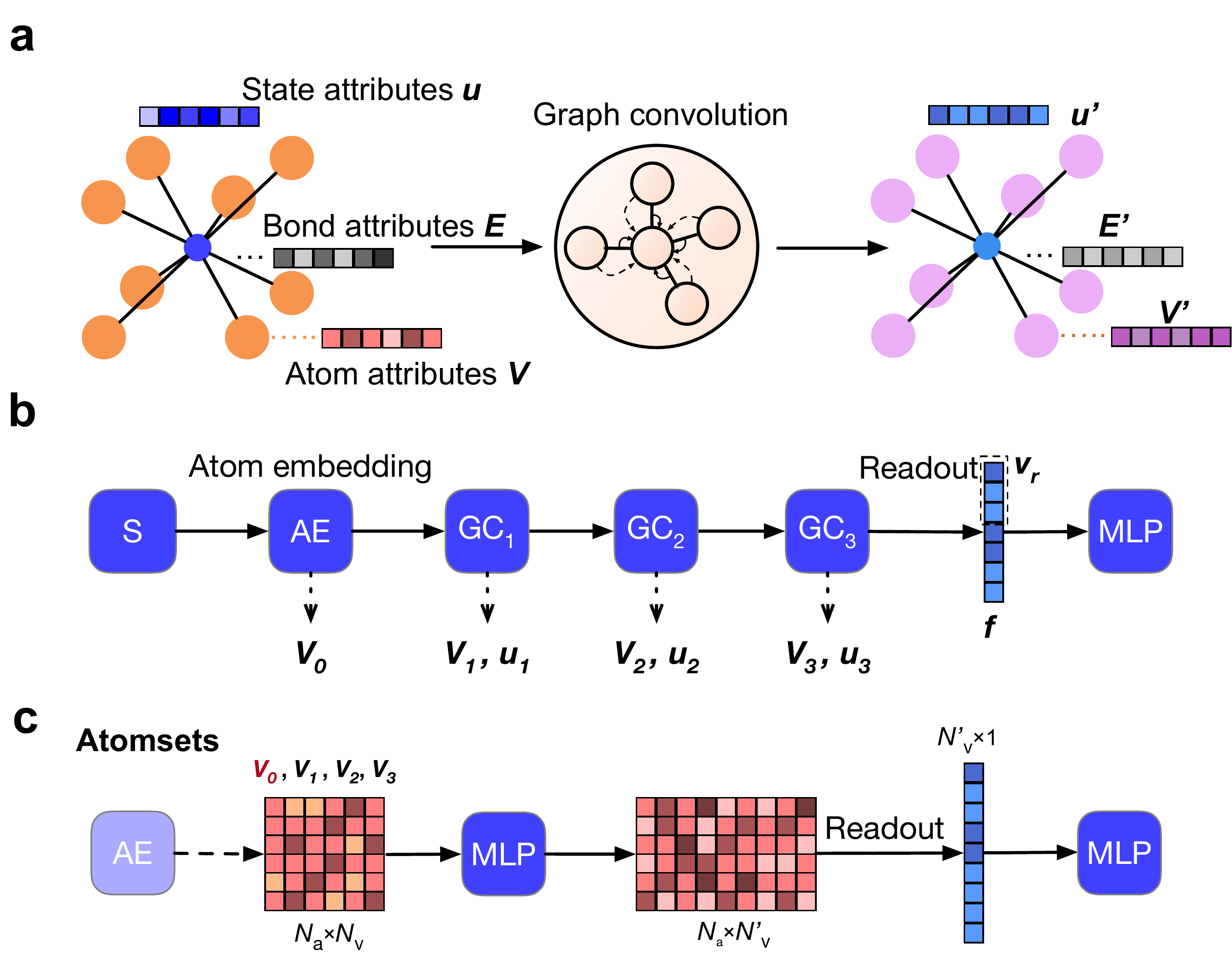}
\caption{Graph networks and AtomSets schematics. \textbf{a,} The graph convolution (GC) takes an input graph with labeled atom ($\mathbf{V}$), bond ($\mathbf{E}$) and state ($\mathbf{u}$) attributes and outputs a new computed graph with updated attributes. \textbf{b,} The graph network model architecture. The input to the model is the structure graph (S) with atomic number as the atom attributes. Then the graph is passed to an atom embedding (AE) layer, followed by three GC layers. After the GC, the graph is readout to a structure-wise vector $f$, and $f$ is further passed to multi-layer perceptron (MLP) models. Within the model, each layer output is captured for later use. \textbf{c,} The AtomSets model takes a site-wise/element-wise feature matrix and passes to MLP layers. After the MLP, the a readout function is applied to derive a structure-wise/formula-wise vector, followed by final MLP layers. }
\label{fig:schematics}
\end{figure}

The MEGNet formalism has been described extensively in previous works\cite{chenGraphNetworksUniversal2019,chenLearningPropertiesOrdered2021} and interested readers are referred to those publications for details. Briefly, the MEGNet framework featurizes a material into a graph $G = (V, E, \mathbf{u})$, where $\mathbf{v}_i \in V$ are the atom or node features, $\mathbf{e}_k \in E$ are the edges or bonds, and $\mathbf{u}$ are state features. The features matrices/vectors are $\mathbf{V} = [\mathbf{v}_1;...; \mathbf{v}_{N_a}] \in \mathbb{R}^{N_a\times N_v}$, $\mathbf{E} = [\mathbf{e}_1;...; \mathbf{e}_{N_b}] \in \mathbb{R}^{N_b\times N_{bf}}$ and $\mathbf{u} \in \mathbb{R}^{N_u}$, where $N_a$, $N_b$, $N_{f}$, $N_{bf}$ and $N_u$ are the number of atoms, bonds, atom features, bond features, state features, respectively. For compositional models, $N_a$ is the number of atoms in the formula. For simplicity, the atom and bond features are represented as matrices. However, shuffling the first dimension does not change the results of the models. Hence, the atoms and bonds are essentially sets.  A graph convolution (GC) operation uses the connectivity of bonds to transform input graph features $(\mathbf{V}, \mathbf{E}, \mathbf{u})$ to output graph features $(\mathbf{V^\prime}, \mathbf{E^\prime}, \mathbf{u^\prime})$, as shown in Figure \ref{fig:schematics}a, by updating the the atom, bond and state features as follows:

\begin{eqnarray}
    \mathbf{e}_k^{(i)} = \phi_e(\mathbf{e}_k^{(i-1)}, \mathbf{v}^{(i-1)}_{s_k}, \mathbf{v}^{(i-1)}_{r_k}, \mathbf{u}^{(i-1)}) \label{eq:gc1}\\
    \mathbf{v}_j^{(i)} = \phi_v(\mathbf{v}_j^{(i-1)}, \mathbf{v}_{k\in \mathcal{N}(j)}^{(i-1)},  
    \mathbf{e}_{k, r_k = j}^{(i)}, \mathbf{u}^{(i-1)}) \label{eq:gc2}\\
    \mathbf{u^{(i)}} = \phi_u(\frac{1}{N_b}\sum_k{\mathbf{e}_k^{(i)}}, \frac{1}{N_a} \sum_k{\mathbf{v}_k^{(i)}}, \mathbf{u}^{(i-1)}) \label{eq:gc3}
\end{eqnarray}

where $i$ is an index indicating the layer of the GC, $\mathbf{e}_k^{(i)}$ and $\mathbf{v}_j^{(i)}$ are bond attributes of bond $k$ and atom attributes of atom $j$ at layer $i$ respectively, $\phi$s are the update functions approximated using multi-layer perceptrons (MLPs), $\mathcal{N}(j)$ indicates the neighbor atom indices of atom $j$, and $r_k = j$ are the bonds with receiving atom index as $j$. 

In the initial structural graph ($S$), the atom attributes are simply the atomic number of the element embedded into a vector space via an atom embedding (AE) layer ($AE: \mathbb{Z} \xrightarrow[]{} \mathbb{R}^{N_f^0}$) to obtain $\mathbf{V_0} \in \mathbb{R}^{N_a\times N_f^0}$, as shown in Figure \ref{fig:schematics}b. The bonds are constructed by considering atom pairs within certain cutoff radius $R_c$. With each GC layer, information is exchanged between atoms, bonds and state. As more GC layers are stacked (e.g., GC$_2$ and GC$_3$ in Figure \ref{fig:schematics}b), information on each atom can be propagated to further distances. 

In this work, a MEGNet model with three GC layers was first trained on the formation energies of more than 130,000 Materials Project crystals as of Jun 1 2019, henceforth referred to as the ``parent'' model. The training procedures and hyperparameter settings of the MEGNet models are similar as the previous work.\cite{chenGraphNetworksUniversal2019}

\subsection{AtomSets Framework}

In our proposed AtomSets framework, the output atom $\mathbf{V}_i = [\mathbf{v}_1^{(i)};...; \mathbf{v}_{N_a}^{(i)}]$ features after each GC layer are extracted from the parent model and transferred to develop models for other properties. Bond features are not considered in TL since the number of bonds depends on the graph construction settings and parameters, such as cutoff radius. As shown in Figure \ref{fig:schematics}c, an AtomSets model takes the atom-wise features $\mathbf{V}_i$ matrix of shape $N_a \times N_f$ as inputs to an MLP model. These features can either be compositional, e.g., elemental properties, or structural, e.g., local environment descriptor. Afterwards, the output feature matrix is readout to a vector, compressing the atom number dimension.

The purpose of the readout function is to reduce the feature matrices with different number of atoms to structure-wise vectors subject to permutational invariance. Simple functions to calculate the statistics along the atom number dimension can be used as readout functions. In this work, we tested two types of readout functions. The linear mean readout function averages the feature vectors, as follows

\begin{equation}
    \bar{\mathbf{x}} = \frac{\sum_i w_i \mathbf{x_i} }{\sum_i w_i} 
\label{eqn:avg}
\end{equation}

where $\mathbf{x_i}$ is the feature row vector for atom $i$ and $w_i$ is the corresponding weights. The weights are atom fractions on one site, e.g., $w_{\rm{Fe}} = 0.01$ and  $w_{\rm{Ni}} = 0.99$ in \ce{Fe_{0.01}Ni_{0.99}}.  We also tested a weight-modified attention-based set2set\cite{vinyalsOrderMattersSequence2016} readout function. We start with memory vectors $\mathbf{m_i}$ = $\mathbf{x_i W} + \mathbf{b}$, and initialize $\mathbf{q_0}^* = \mathbf{0}$, where $\mathbf{W}$ and $\mathbf{b}$ are learnable weights and biases respectively. At step $t$, the updates are calculated using long short-term memory (LSTM) and attention mechanisms as follows

\begin{eqnarray}
\mathbf{q_t} = LSTM(\mathbf{q_{t-1}^*}) \label{eqn:start_set2set} \\
e_{i, t} = \mathbf{m_i}\cdot \mathbf{q_t} \\
a_{i, t} = \frac{w_i\exp{(e_{i, t})}}{\sum_j w_j\exp{(e_{j, t})}} \\
\mathbf{r_t} = \sum_i{a_{i, t}\mathbf{m_i}} \\
\mathbf{q_t^*} = \mathbf{q_t}\oplus \mathbf{r_t}
\label{eqn:set2set}
\end{eqnarray}
A total of three steps are used in the weighted-set2set readout function.
 
Then, the readout vector can be used to predict properties with the help of MLP or other models, as shown in Figure \ref{fig:schematics}c. The feature matrix can either be taken as pre-trained model generated feature matrices $\mathbf{V_i}$ ($i=0, 1, 2, 3$) or trained on-the-fly via a trainable atom embedding layer prepended to the model.

When the site-wise/atom-wise features are computed from pre-trained models, information gained from previous model training is retained and effectively the AtomSets models transfer-learn part of the pre-trained models. A hierarchical TL scheme is achieved by including different GC outputs.  The AtomSets models can also be used without transfer learning, by training the elemental embedding and hence atom-wise features from the data. 

The AtomSets framework is flexible in the choice of input features. For example, if the symmetry functions are provided as inputs, then the AtomSets model becomes the high-dimensional neural network potential.\cite{behlerGeneralizedNeuralNetworkRepresentation2007} The AtomSets framework also shares similarity with the Deepsets\cite{zaheerDeepSets2017} model where the summation of feature vectors are used to get the readout vectors. Since only simple MLP are underlying the AtomSets framework, the model training can be extremely fast. Models investigated in the current work are provided in Table \ref{tab:model_type}. 


\begin{table}[h!]
\caption{\label{tab:model_type} Models investigated in this work, categorized by the models types, i.e., compositional (C) or structural (S), and whether they utilize transfer learning (TL). In our definition, S-type models contain compositional information as a superset. It should be noted that the MLP-$u_i$, MLP-$f$ and MLP-$v_r$ are classified as S-type models because $u_i$, $f$ or $v_r$ implicitly incorporate structural information due to information passing in the graph convolution layers.}
\begin{center}
\begin{tabular}{  L{4cm} | C{1cm}| C{1cm} | L{8cm}  } 
\hline
Model name & Type & TL & Description \\
 \hline
AtomSets & C & No & Compositional models directly trained from data\\
\hline
AtomSets-$V_0$ & C & Yes & Compositional models transferring learned $V_0$ from the parent formation energy model\\
\hline
AtomSets-$V_i$ ($i=1, 2, 3$) & S & Yes & Structural models transferring learned $V_1$, $V_2$ or $V_3$ features from the parent formation energy model\\
\hline
MLP-$V_0$-stats & C & Yes & Compositional MLP models using statistics calculated on $V_0$ from the parent formation energy model as inputs \\
\hline
MLP-$u_i$ ($i=1, 2, 3$), MLP-$f$ and MLP-$v_r$ & S & Yes & MLP models using learned $u_i$, $f$ or $v_r$ from the parent formation energy model.\\
\hline
MEGNet & S & No & Graph network models trained directly using each data set without transfer learning\\
\hline

\end{tabular}
\end{center}
\end{table}

\subsection{Data and Model Training}

The 13 materials datasets were obtained from the matbench repository\cite{dunnBenchmarkingMaterialsProperty2020}. A summary is provided in Table \ref{tab:data}, where the data sizes range from 312 to 132,752, with both compositional and structural data. The tasks include regression and classification. Detailed summaries are provided in the work by \citet{dunnBenchmarkingMaterialsProperty2020} For each model training, we split the data into 80\%-10\%-10\% train-validation-test sets. The validation set was used to stop the model fitting when the validation metric, i.e., mean-absolute-error (MAE) in regression and area under the curve (AUC) in classification, did not improve for more than 200 consecutive epochs. The model with the lowest validation error was chosen as the ``best'' one. Each model was fitted five times using different random splits, and the average and standard deviations of the metric on the test set were reported. 

A grid search was performed on the hyperparameters for the AtomSets models and MLP models. The parameter candidates are provided in Table \ref{tab:hyperparameters}. During the screening process, a 5-fold random shuffle split is applied to the data set, and the parameter set with the lowest average validation error was chosen. The \textit{matbench\_steels} (compositional) and \textit{matbench\_phonons} (structural) data sets were first used to perform an initial screening for relatively good parameter sets. Starting from these parameter sets, a further grid search for all datasets for the most generalizable AtomSets-$V_0$ (compositional) and AtomSets-$V_1$ (structural) models was then performed.

\section{Results}

\subsection{Model accuracies}

The MAE of regression and the AUC of classification for various tasks are shown in Table \ref{tab:benchmark}. In addition, hyperparameter optimization was carried out on the AtomSets-V$_0$ and V$_1$ models (see Table \ref{tab:best_models}), but did not seem to have a significant effect on model performance. Here, we will focus our discussion on the models without further hyperparameter optimization. To frame our analysis, we will first recapitulate that a key finding of \citet{dunnBenchmarkingMaterialsProperty2020} is that MEGNet models tend to outperform other models when the data size is large ($> 10,000$ data points) but underperform for small data sizes. This can be seen in the last two columns of Table \ref{tab:benchmark}, where AutoMatminer models achieve lower MAEs for the small yield strength, exfoliation energies and refractive index data sets compared to MEGNet. 

\begin{table}[h!]
\begin{center}
\begin{adjustbox}{max width=\textwidth}
\caption{Performance of AtomSets models relative to state-of-the-art models. The average and standard deviations of the MAE and AUC are reported for regression and classifcation tasks, respectively. The properties are sorted by dataset size. Some structural models (e.g., AtomSets-$V_1/V_2/V_3$ for experimental band gaps) cannot be constructed as the dataset does not contain structural information. The best performing model(s) within the standard deviation are bolded for each target.} 
\label{tab:benchmark}
\begin{tabular}{  L{5cm} |C{1.8cm}|C{1.8cm} | C{1.8cm} | C{1.8cm} | C{1.8cm} | C{1.8cm} | C{1cm} | C{2cm}} 
\hline
\hline
Target, Data Size & AtomSets & AtomSets-$V_0$ & AtomSets-$V_1$ & AtomSets-$V_2$ & AtomSets-$V_3$ & MLP-$f$ & MEGNet\cite{dunnBenchmarkingMaterialsProperty2020} & AutoMatminer\cite{dunnBenchmarkingMaterialsProperty2020}\\
\hline
\hline
\multicolumn{9}{c}{Regression Tasks}\\
\hline
Yield Strength (GPa), 312$^a$ & 109$\pm$14 &\textbf{90$\pm$25} & -  & - & - & - & - & \textbf{95}\\
\hline
$E_{exfoliation}$ (meV/atom), 636$^b$ & 52$\pm$7& 51$\pm$6 & \textbf{40$\pm$5} & 48$\pm$9 & 54$\pm$7 & 55$\pm$2 & 56 & \textbf{39}\\
\hline
PhonDOS Peak (1/cm), 1265$^c$ & 63$\pm$6& 52$\pm$7 & 54$\pm$9 & 106$\pm$4 & 126$\pm$75 & 154$\pm$8 & \textbf{37} & 51\\
\hline
Expt. $E_g$ (eV), 4604$^d$ &0.45$\pm$0.02  & \textbf{0.43$\pm$0.03} & - & - & - & - & - & \textbf{0.42}\\
\hline
$\varepsilon$, 4764$^e$& 0.51$\pm$0.05& 0.47$\pm$0.05 & 0.46$\pm$0.03 & 0.52$\pm$0.08 & 0.56$\pm$0.06 & 0.59$\pm$0.05 & 0.48 & \textbf{0.30 }\\
\hline
log($K_{VRH}$) (GPa), 10987$^f$& 0.09$\pm$0.00 & 0.09$\pm$0.00 & \textbf{0.07$\pm$0.00} & 0.09$\pm$0.00 & 0.15$\pm$0.07 & 0.10$\pm$0.00 & \textbf{0.07} & \textbf{0.07}\\
\hline
log($G_{VRH}$) (GPa), 10987$^g$& 0.11$\pm$0.00& 0.11$\pm$0.00 & \textbf{0.09$\pm$0.00} & 0.10$\pm$0.00 & 0.14$\pm$0.08 & 0.11$\pm$0.00 & \textbf{0.09} & \textbf{0.09}\\
\hline
Perovskite $E_f$ (meV/atom), 18928$^h$& 83$\pm$1 & 83$\pm$1 & 12$\pm$0 & 24$\pm$0 & 113$\pm$1 & 30$\pm$0 & \textbf{8} & 39\\
\hline
MP $E_g$ (eV), 106113$^i$& 0.26$\pm$0.00&0.27$\pm$0.00 & \textbf{0.25$\pm$0.01} & 0.30$\pm$0.01 & 0.30$\pm$0.01 & 0.37$\pm$0.00 & \textbf{0.24} & 0.28\\
\hline
MP $E_f$ (meV/atom), 132752$^j$ & 107$\pm$2& 108$\pm$1 & 44$\pm$1 & 66$\pm$0 & 73$\pm$1 & \textbf{17$\pm$0} & 33 & 173\\\hline
\hline
\multicolumn{9}{c}{Classification Tasks}\\
\hline
Expt. Metallicity, 4921$^k$& \textbf{0.94$\pm$0.02}& \textbf{0.94$\pm$0.01} & - & - & - & - & - & 0.92\\
\hline
Glass Forming Ability, 5680$^l$ & \textbf{0.92$\pm$0.02} & \textbf{0.92$\pm$0.01} & -  & - & - & - & - & 0.86\\
\hline
MP Metallicity, 106113$^m$& 0.95$\pm$0.00&0.95$\pm$0.00 & 0.96$\pm$0.00 &0.95$\pm$0.00 & 0.96$\pm$0.00 & 0.96$\pm$0.00 & \textbf{0.98} & 0.91 \\
\hline
\hline
\end{tabular}
\end{adjustbox}
\end{center}

\begin{flushleft}
\small{$^a$Steel yield strength data from Citrine Informatics.\cite{conduitCitrination2017}\\ $^b$Exfoliation energy of crystals from JARVIS DFT 2D dataset.\cite{choudharyHighthroughputIdentificationCharacterization2017}\\
$^c$Phonon DOS peak frequency from Materials Project.\cite{petrettoHighthroughputDensityfunctionalPerturbation2018}\\
$^d$Experimental composition-band gap dataset from \citet{zhuoPredictingBandGaps2018}\\
$^e$Refractive index from Materials Project.\cite{petousisHighthroughputScreeningInorganic2017}\\
$^f$Log of computed bulk moduli from Materials Project.\cite{dejongChartingCompleteElastic2015}\\
$^g$Log of computed shear moduli from Materials Project.\cite{dejongChartingCompleteElastic2015}\\
$^h$Computed perovskite formation energy from \citet{castelliNewCubicPerovskites2012}\\
$^i$Computed PBE band gap data from Materials Project\cite{jainCommentaryMaterialsProject2013,ongMaterialsApplicationProgramming2015}.\\
$^j$Computed PBE formation energy data from Materials Project\cite{jainCommentaryMaterialsProject2013,ongMaterialsApplicationProgramming2015}.\\ $^k$Experimental metallicity (binary) from \citet{zhuoPredictingBandGaps2018}\\
$^l$Glass forming ability (binary) from Landolt-Bornstein Handbook\cite{kawazoeNonequilibriumPhaseDiagrams1997}.\\
$^m$Computed PBE metallicity (binary) from Materials Project.\cite{jainCommentaryMaterialsProject2013,ongMaterialsApplicationProgramming2015}}
\end{flushleft}
\end{table}

The AtomSets models do not suffer from the same data size tradeoff observed in the MEGNet models. With a few notable exceptions, the transfer-learned AtomSets models usually achieve close to the best performance (lowest MAE or highest AUC, with the error bar) among all models studied. For the very small yield strength and $E_{exfoliation}$ datasets, AtomSets models perform similarly to AutoMatminer, while for the larger formation energies (Perovskite and MP $E_f$) and MP band gap ($E_g$) datasets, AtomSets models perform similarly to MEGNet. The only dataset where the AtomSets and MEGNet models substantially underperform relative to AutoMatminer is the refractive index of crystals from the Materials Project. This suggests that some of the additional features considered in the AutoMatminer algorithm (e.g., electronic structure of the constituent elements) might be necessary for ML algorithms to predict the refractive index.


A somewhat surprising observation is that several target properties show minimal dependency on structural information. For example, the average MAEs of the compositional AtomSets-$V_0$ models and structural AtomSets-$V_1$ models for the JDFT-2D exfoliation energy, the MP phonon DOS peak, and the refractive index datasets are within the standard deviation. The structural AtomSets-$V_1$models for the MP elasticity data ($\log K_{VRH}$ and $\log G_{VRH}$) only exhibit minor improvements in average MAEs over the compositional AtomSets-$V_0$ models. To investigate the implications of this observation, we analyzed the polymorphs for each composition in the elasticity data set, see Figure \ref{fig:poly_counts}. Out of the 10987 elasticity data, 81\% of them do not have polymorphs. For those materials, structural models likely perform similarly to the compositional models. For compositions that have more than one polymorph (816 out of 9723 unique compositions), we calculated the range of the target values for polymorphs, as shown in Figure \ref{fig:poly_counts}b and c. The majority of the polymorphs for the same composition have similar bulk and shear moduli, and the average ranges for $\log K_{VRH}$ and $\log G_{VRH}$ are 0.134 and 0.158, respectively. If we include compositions with no polymorphs, i.e., range equals zero, the average ranges for $\log K_{VRH}$ and $\log G_{VRH}$ are 0.011 and 0.013, respectively. Such small ranges for each composition suggest that composition explains the majority of the variation in bulk moduli, which is why the accuracy differences between AtomSets-$V_0$ and AtomSets-$V_1$ are minimal. This observation also gives a glance at why compositional models have been reasonably successful. It should be noted that there are well-known polymorphs with vastly different mechanical properties, e.g., diamond and graphite carbon, and the AtomSets-$V_1$ provide far better predictions. For example, the AtomSets-$V_1$ model predicts the shear moduli of graphite (96 GPa) and diamond (520 GPa) to be 96 GPa and 490 GPa, respectively, while the AtomSets-$V_0$ model predicts them to be 177 GPa.  In contrast, the perovskites and MP formation energy datasets require structural models to achieve accurate results. This observation is consistent with a recent study by \citet{bartelCriticalExaminationCompound2020}

Comparing AtomSets models with various $\mathbf{V}$'s, the results show that the features extracted from earlier stage GC layers, e.g., $\mathbf{V_0}$ and $\mathbf{V_1}$, are more generalizable and have higher accuracy in all models compared to those produced by later GC layers. The structure-wise state vectors, $\mathbf{u_i}~(i = 1, 2, 3)$, and the readout atom feature vector $\mathbf{v_r}$, are relatively poor features, as shown by the large errors in all models in Table \ref{tab:benchmark_si}. However, the final structure-wise readout vector $\mathbf{f}$, along with MLP models, offers excellent accuracy in MP metallicity and formation energy tasks. 

\subsection{Model Convergence}

\begin{figure}
\centering
\includegraphics[width=0.9\textwidth]{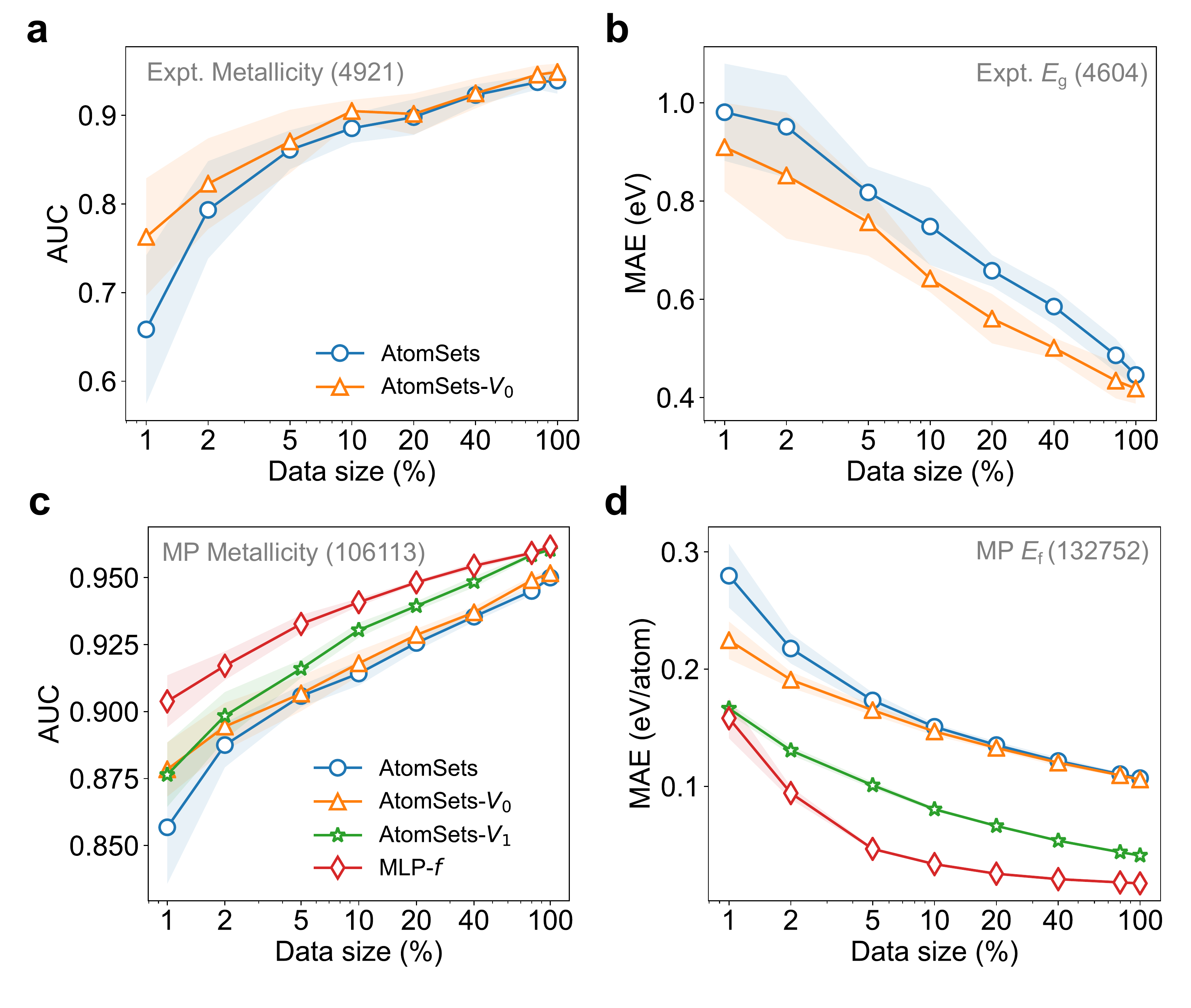}
\caption{Model convergence for AtomSets, AtomSets-$V_0$, AtomSets-$V_1$ and MLP-$f$ of small compositional (a and b) and large structural (c and d) datasets. (a) and (c) show the area under the curve (AUC) for classification tasks,  and (b) and (d) show the mean absolute error (MAE) for regression tasks. The x-axis is plotted on a log scale to provide improve resolution at small data sizes. The shaded areas are the standard deviation across five randomly data fitting. Additional model results are shown in Figure \ref{fig:bench_conv}}
\label{fig:convergence}
\end{figure}

A convergence study of the best models - two compositional models, i.e., AtomSets, AtomSets-$V_0$, and two structural models, i.e., AtomSets-$V_1$ and MLP-$f$ - was performed relative to data size. Different data sizes in terms of the fractions of maximum available data are applied. Comparing the two compositional models, the AtomSets-$V_0$ model achieves relatively higher performance throughout all the tasks and generally converges faster than the non-TL counterpart, i.e., the AtomSets model, as shown in Figure \ref{fig:convergence}. For the structural datasets in Figure \ref{fig:convergence}c and  \ref{fig:convergence}d, consistent with previous benchmark results, the structural AtomSets-$V_1$ and MLP-$f$ models are generally more accurate than the compositional models. The rapid convergence of the MLP-$f$ models in the MP formation energy dataset is expected since the structural features $\mathbf{f}$ were generated by the formation energy MEGNet models in the first place. Model convergences on the rest of the datasets are provided in Figure \ref{fig:bench_conv}.

\begin{figure}
\centering
\includegraphics[width=0.9\textwidth]{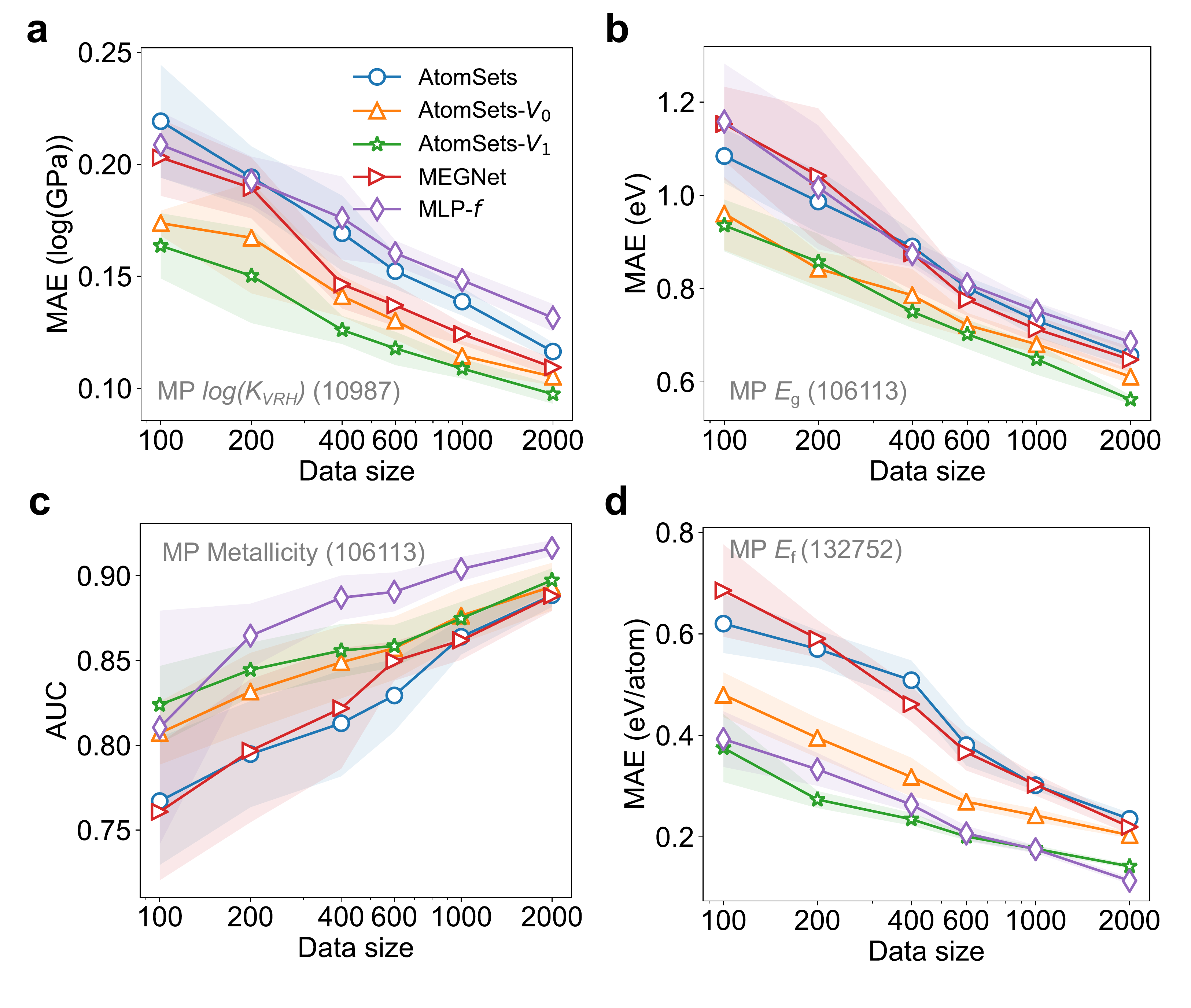}
\caption{Model convergence in the small data limits. The four datasets are the (a) log10 of the bulk moduli, (b) band gap, (c) binary metallicity and (d) formation energy structural datasets from the Materials Project.}
\label{fig:small_convergence}
\end{figure}

The model performance is also probed at tiny datasets. We used several MP datasets in this study to obtain consistent results and then down-sampled the datasets at 100, 200, 400, 600, 1000, and 2000 data points. For comparison, we also include the non-TL MEGNet structural models, as shown in Figure \ref{fig:small_convergence}. Similar to the previous convergence study at relatively large data sizes, the TL compositional models AtomSets-$V_0$ outperform the non-TL compositional AtomSets models at all data sizes. For structural models, the TL AtomSets-$V_1$ models achieve consistent accuracy at small data limits for all four tasks and consistently outperform the non-TL MEGNet models.

Interestingly, the MLP-$f$ models specialize in MP metallicity data and MP formation energy data, same as previous benchmark results shown in Table \ref{tab:benchmark}. In particular, the MLP-$f$ models converge rapidly for the MP metallicity task, with AUC exceeding 85\% with only 200 data points and 90\% with only 1000 data points. The MLP-$f$ models also reach $\sim$ 0.2 eV/atom errors on the MP formation energy data when the data size is 600. In both cases, the MLP-$f$ models outperform MEGNet models by a considerable margin. However, in terms of generalizability, the AtomSets-$V_1$ models seem to be a better fit for all generic tasks. 

At a data size of 600 (533 train data points), the formation energy and the band gap models errors of AtomSets-$V_1$ are 0.2 eV/atom and 0.702 eV, respectively, much lower than the errors achieved by the full MEGNet models with 0.367 eV/atom and 0.78 eV. The AtomSets-$V_1$ errors at such small data regimes are on par with the 0.210 eV/atom and 0.71 eV errors (504 train data points) reported by the MODNet models\cite{debreuckMachineLearningMaterials2020} that specialize in small materials data fitting. Interestingly, the compositional model AtomSets-$V_0$ also achieved lower errors than full MEGNet, with formation energy model errors of 0.269 eV/atom and band gap model errors of 0.72 eV.

\subsection{Model Extrapolability}
\begin{figure}
\centering
\includegraphics[width=0.9\textwidth]{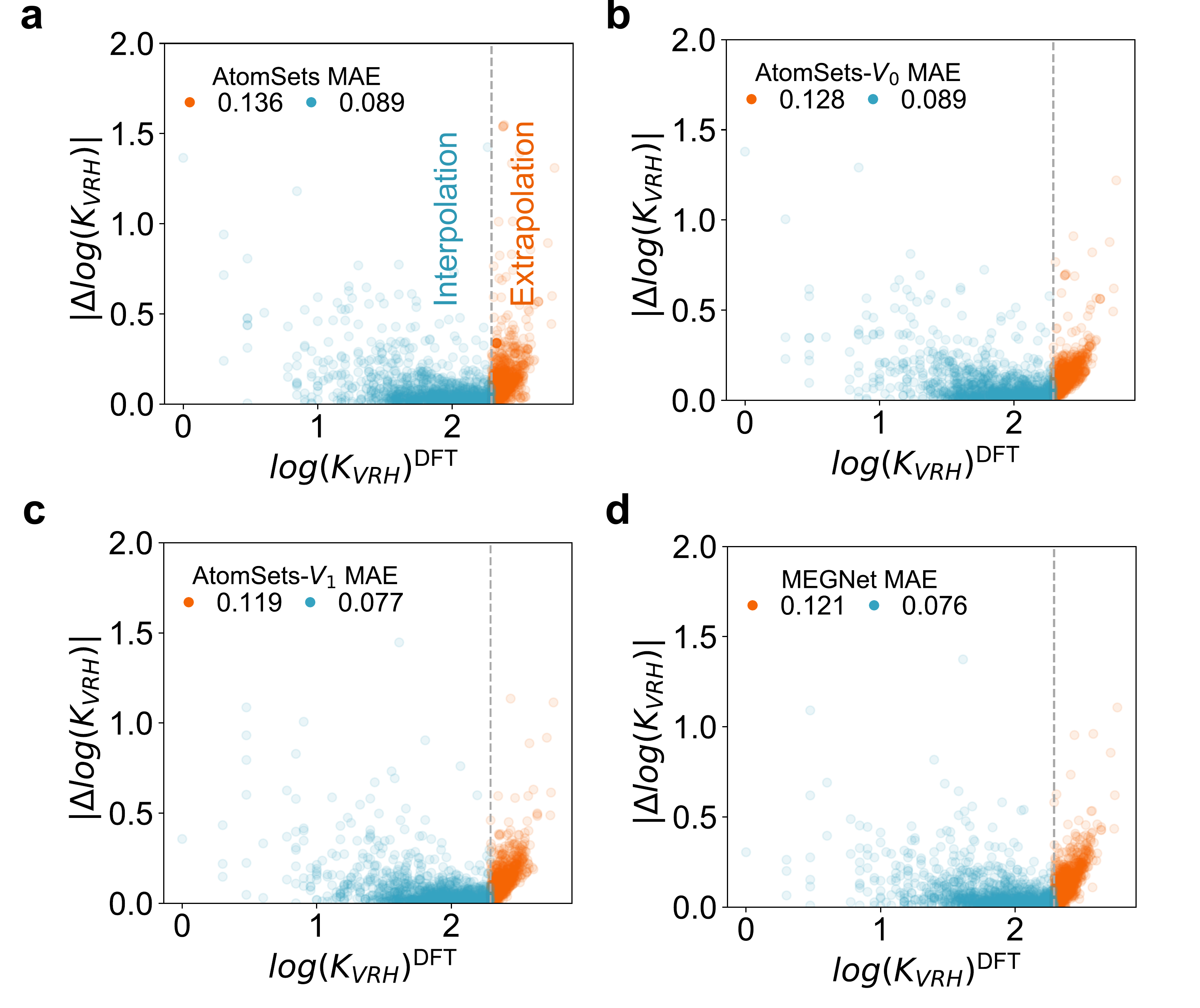}
\caption{Absolute differences in predicted and DFT $\log(K_{VRH})$, i.e., $|\Delta \log(K_{VRH})|$ against the DFT value range for the test data. The training and validation data are randomly sampled from the 0\% to 90\% (vertical dash line) target quantile range. Half of test data comes from the 90\%-100\% quantile (extrapolation) and the other half is from the same target range as the train-validation data (interpolation).}
\label{fig:extrapolation}
\end{figure}
In a typical materials design problem, the target is not finding a material with similar performance as most existing materials, but rather materials with extraordinary properties that lie outside of the current materials pool. Such extrapolation presents a major challenge for most ML models. Previous works have designed leave-one cluster out cross-validation (LOCO CV)\cite{meredigCanMachineLearning2018} or k-fold forward cross-validation\cite{xiongEvaluatingExplorativePrediction2020} to evaluate the models' extrapolation ability in data regions outside the training data. Here we adopted the concept of forward cross-validation by splitting the data according to their target value ranges and applied the method to elasticity data (MP $log(K_{VRH})$ and MP $log(G_{VRH})$) to imitate the process of finding super-incompressible (high $K$) and superhard (roughly high $G$) materials. First, we held out the materials with the top 10\% corresponding target values as the test dataset (high-test, extrapolation). Then for the remaining, we also split it into the train, validation, and test (low-test, interpolation) datasets, making two test data regimes in total. We selected AtomSets, AtomSets-$V_0$, AtomSets-$V_1$, and the MEGNet models for the comparison. For the bulk moduli $K$, the low-test errors for the compositional models AtomSets and AtomSets-$V_0$ are identical. However, with the test target value outside of the training data range, the errors increase rapidly above the low-test errors. Nevertheless, the TL model AtomSets-$V_0$ are better generalized in the extrapolation high-test regime, as shown by the lower extrapolation errors in Figure \ref{fig:extrapolation}a and Figure \ref{fig:extrapolation}b. For structural models, the low-test errors are again almost the same, yet the TL AtomSets-$V_1$ models have lower errors than the MEGNet counterparts, see Figure \ref{fig:extrapolation}c and Figure \ref{fig:extrapolation}d. 
Similar conclusions can be reached using the shear moduli dataset, as shown in Figure \ref{fig:logg}. These results conclude that TL approaches can significantly enhance the models' accuracy in extrapolation tasks critical in new materials discovery.

\section{Discussion}
The hierarchical MEGNet features provide a cascade of descriptors that capture both short-ranged  interactions at early GC (e.g., $\mathbf{V_0}, \mathbf{V_1}$) and long-ranged interactions at later GC (e.g., $\mathbf{V_2}, \mathbf{V_3}$). The first GC features are better TL features across various tasks, while the latter GC generated features generally exhibit worse performance.  We can explain this part by drawing an analogy to convolutional neural networks (CNN) in facial recognition, where the early feature maps capture generic features such as lines and shapes and the later feature maps form human faces.\cite{leeConvolutionalDeepBelief2009} It is not surprising that if such CNN is transferred to other domains, for example, recognizing general objects beyond faces, the early feature maps may work, while the later ones will not. 

One surprising result from our studies is the relatively good performance of the compositional models (AtomSets-$V_0$) on many properties, e.g., the phonon dos and bulk and shear moduli. It would be erroneous to conclude that these properties are not structure-dependent. We believe the main reason for the outperformance of the compositional models is because most compositions either do not exhibit polymorphism or have many polymorphs with somewhat similar properties, e.g., the well-known family of SiC polymorphs. These results highlight the importance of generating a diversity of data beyond existing known materials. Existing databases such as the Materials Project typically prioritize computations on known materials, e.g., ICSD crystals. While such a strategy undoubtedly provides the most value to the community for the study of existing materials, the discovery of new materials with extraordinary properties require exploration beyond known materials; additional training data on hypothetical materials is critical for the development of ML models that can extrapolate beyond known materials design spaces. The use of TL, as shown in this work, is nevertheless critical for improving the extrapolability of models. 

The AtomSets framework can be viewed as a particular case of the graph network models\cite{battagliaRelationalInductiveBiases2018, chenGraphNetworksUniversal2019} without the edge and global information update. With TL, the previously learned edge and global information from MEGNet model training are implicitly included to the node information and hence the AtomSets model. The AtomSets greatly simplify the graph network models and thus can be trained at a much small computational cost without compromising the model accuracy. For example, it takes about 10 seconds per epoch for training AtomSets model on the most extensive MP formation energy data (132,752) using one GTX 1080Ti GPU, while training MEGNet can take $>300$ seconds/epoch. 

\section{Conclusion}

This work introduces a new and straightforward deep learning model framework, the AtomSets, as an effective way to learn materials properties at all data sizes and for both compositional and structural data. By combining with TL, the structure-embedded compositional and structural information can be readily incorporated into the model. The simple model architecture makes it possible to train the models with much smaller datasets and lower computational resources compared to graph models. We show that the AtomSets models can achieve consistently low errors for small data tasks, e.g., steel strength datasets, to extensive data tasks, e.g., MP computational data, and the model accuracy further improves with TL. We also show better model convergence for the AtomSets models. The AtomSets framework introduces a facile deep learning framework and helps accelerate the materials discovery process by combining accurate compositional and structural materials models. 

\section{Data Availability}
The MatBench datasets are available from the AutoMatminer\cite{dunnBenchmarkingMaterialsProperty2020} github repository (\url{https://github.com/hackingmaterials/automatminer}).

\section{Code Availability}
The AtomSets framework and MEGNet featurizations are implemented in the open source materials machine learning (maml) package\cite{chenMamlMaterialsMachine2021} (\url{https://github.com/materialsvirtuallab/maml}).

\begin{acknowledgement}
The authors acknowledge the support from the Materials Project, funded by the U.S. Department of Energy, Office of Science, Office of Basic Energy Sciences, Materials Sciences and Engineering Division under contract no. DE-AC02-05-CH11231: Materials Project program, KC23MP. The authors also acknowledge computational resources provided by the Triton Shared Computing Cluster (TSCC) at the University of California, San Diego, and the Extreme Science and Engineering Discovery Environment (XSEDE) supported by the National Science Foundation under grant no. ACI-1053575.

\end{acknowledgement}

\section{Author contributions}
C.C. and S.P.O. conceived the idea. C.C. carried out the model construction, and fitting under the supervision of S.P.O. C.C. and S.P.O. wrote the manuscript. 

\section{Declaration of interests}

The authors declare no competing interests.

\textbf{Supplementary Information} accompanies this paper at 

\clearpage
\bibliography{refs}

\end{document}


\maketitle
\newpage

\begin{table}[h!]
\begin{center}
\caption{Materials data name, data sizes, input types, property name, units and task types. Type shows the input data type, where Comp means composition and Struct means structure. The task includes regression (R) and classification (C). For the matbench\_perovskites datasets, the original data shows formation in eV, while in the final presentation of error metrics, we converted it into eV/atom.} 
\label{tab:data}
\begin{tabular}{  C{5cm} | C{1.5cm}| C{2cm} | C{4cm} | C{2cm} | C{1cm} } 
\hline
Data name & Size & Type & Target & Unit & Task\\ 
\hline
matbench\_steels & 312 & Comp & Yield strength & GPa & R \\ 
\hline
matbench\_jdft2d & 636 & Struct & Exfoliation energy & meV/atom & R\\
\hline
matbench\_phonons & 1265 & Struct & Peak frequency & 1/cm & R \\
\hline
matbench\_dielectric & 4764 & Struct & Dielectric constant & - & R \\
\hline
matbench\_expt\_gap & 4604 & Comp & Band gap & eV & R \\
\hline
matbench\_expt\_is\_metal & 4921 & Comp & Metallicity & - & C \\ 
\hline
matbench\_glass & 5690 & Comp & Metallicity & - & C \\
\hline 
matbench\_log\_kvrh & 10987 & Struct & Bulk moduli & log(GPa) & R \\
\hline
matbench\_log\_gvrh & 10987 & Struct & Shear moduli & log(GPa) & R \\
\hline
matbench\_perovskites & 18928 & Struct & Formation energy & eV & R \\
\hline
matbench\_mp\_gap & 106113 & Struct & PBE band gap & eV &  R\\
\hline
matbench\_mp\_is\_metal & 106113 & Struct & PBE metallicity & - &  C\\
\hline
matbench\_mp\_e\_form & 132752 & Struct & Formation energy & eV &  R\\
\hline
\end{tabular}
\end{center}
\end{table}

\begin{table}[h!]
\begin{center}
\caption{Hyperparameters for AtomSets and MLP. The AtomSets model has two MLPs as components: one before the readout, and one after the readout.} 
\label{tab:hyperparameters}
\begin{tabular}{  C{5cm} | C{10cm} } 
\hline
Parameter name & Possible values \\
\hline
\multicolumn{2}{c}{\textbf{AtomSets}} \\
\hline
n\_layer\_before\_readout & 1, 2, 3 \\
\hline
n\_layer\_after\_readout & 2, 3 \\
\hline
n\_neurons\_per\_layer & 16, 32, 64, 128 \\
\hline
readout\_function & ``mean", ``set2set" \\
\hline
activation\_function & ``softplus", ``relu" \\
\hline
\multicolumn{2}{c}{\textbf{MLP}} \\
\hline
n\_layer & 1, 2, 3, 4, 5 \\
\hline
n\_neurons\_per\_layer & 16, 32, 64, 128 \\
\hline
activation\_function & ``softplus", ``relu" \\

\hline
\end{tabular}
\end{center}
\end{table}

\begin{table}[h!]
\begin{center}
\caption{AtomSets model performance with best parameters.} 
\label{tab:best_models}
\begin{tabular}{  L{8cm} | C{3cm} | C{3cm} } 
\hline
Target, Data Size & AtomSets-$V_0^+$ & AtomSets-$V_1^+$ \\
\hline
Yield Strength (GPa), 312 & $90\pm25$ & - \\
\hline
Exforliation Energy (meV/atom), 636 & $42\pm8$ & $40\pm5$ \\
\hline
Phonon DOS Peak (1/cm), 1265 & 48$\pm$5 & 54$\pm$9 \\
\hline
Expt. Band Gap (eV), 4604 & 0.42$\pm$0.03 & - \\
\hline
Dielectric constants, 4764 & 0.46$\pm$0.05 & 0.44$\pm$0.04 \\
\hline
Expt. Metallicity, 4921 & 0.95$\pm$0.01 & - \\
\hline
Glass Forming Ability, 5680 & 0.92$\pm$0.01 & - \\
\hline
MP log($K_{VRH}$) (GPa), 10987 & 0.08$\pm$0.00 & 0.07$\pm$0.00\\
\hline
MP log($G_{VRH}$) (GPa), 10987 & 0.11$\pm$0.00 & 0.08$\pm$0.00\\
\hline
Perov. Form. Energy (meV/atom), 18928 & 83$\pm$1 & 10$\pm$0 \\
\hline
MP Band Gap (eV), 106113 & 0.26$\pm$0.00 & 0.23$\pm$0.01 \\

\hline
MP Metallicity, 106113 & 0.95$\pm$0.00 & 0.96$\pm$0.00 \\
\hline
MP Form. Energy (meV/atom), 132752 & 106$\pm$1 & 41$\pm$1 \\
\hline

\hline
\end{tabular}
\end{center}
\end{table}

\begin{table}[h!]
\begin{center}
\begin{adjustbox}{max width=\textwidth}
\caption{Various MLP model performances compared to the dummy models. For regression tasks, MAE is reported. For classification tasks (expt. metallicity, glass forming ability, MP metallicity), the area under curve (AUC) is reported. The  models  without  values  in  some  fields  are structure  models  and  the  ones  with  all  properties  fitted  are  compositional  models. For the dummy models, the metrics are obtained by predicting all mean values for the regression tasks and the majority labels for the classification tasks.} 
\label{tab:benchmark_si}
\begin{tabular}{  L{5cm} |C{1.8cm}|C{1.8cm} | C{1.8cm} | C{1.8cm} | C{1.8cm} | C{1.8cm}} 
\hline
\hline
Target, Data Size & MLP-$V_0$-stats & MLP-$u_1$ & MLP-$u_2$ & MLP-$u_3$ & MLP-$v_r$ & Dummy Models\\
\hline
\hline
Yield Strength (GPa), 312 & 107$\pm$28 & - & - & - & - & 230 \\
\hline
Exfol. Energy (meV/atom), 636 & 61$\pm$8 & 47$\pm$5 & 57$\pm$6 & 62$\pm$10 & 63$\pm$9 & 67\\
\hline
PhonDOS Peak (1/cm), 1265 & 111$\pm$10 & 120$\pm$13 & 224$\pm$13 & 233$\pm$34 & 188$\pm$15 & 324 \\
\hline
Expt. Band Gap (eV), 4604 & 0.61$\pm$0.03 & - & - & - & - & 1.14\\
\hline
Dielectric constants, 4764& 0.59$\pm$0.03 & 0.56$\pm$0.03 & 0.65$\pm$0.05 & 0.67$\pm$0.06 & 0.55$\pm$0.03 & 0.81 \\
\hline
Expt. Metallicity, 4921& 0.95$\pm$0.01 & - & - & - & - & 0.50\\
\hline
Glass Forming Ability, 5680 & 0.92$\pm$0.01  & - & - & - & - & 0.50\\
\hline
log($K_{VRH}$) (GPa), 10987& 0.11$\pm$0.00 & 0.10$\pm$0.00 & 0.15$\pm$0.00 & 0.18$\pm$0.01 & 0.12$\pm$ 0.00 & 0.29\\
\hline
log($G_{VRH}$) (GPa), 10987& 0.13$\pm$0.01 & 0.12$\pm$0.00 & 0.16$\pm$0.01 & 0.18$\pm$0.01 & 0.13$\pm$0.00 & 0.29\\
\hline
Perov. Form. Energy (meV/atom), 18928& 87$\pm$2 & 40$\pm$1 & 51$\pm$1 & 43$\pm$0 & 39$\pm$1 & 113\\
\hline
MP Band Gap (eV), 106113& 0.46$\pm$0.01 & 0.57$\pm$0.01 & 0.60$\pm$0.01 & 0.57$\pm$0.01 & 0.44$\pm$0.00 & 1.33\\

\hline
MP Metallicity, 106113& 0.94$\pm$0.00 & 0.94$\pm$0.00 & 0.93$\pm$0.00 & 0.93$\pm$0.00 &  0.95$\pm$0.00 & 0.50\\
\hline
MP Form. Energy (meV/atom), 132752 & 157$\pm$2 & 125$\pm$1 & 180$\pm$1 & 138$\pm$2 & 79$\pm$1 & 1010\\
\hline

\hline
\end{tabular}
\end{adjustbox}
\end{center}
\end{table}

\begin{figure}
\centering
\includegraphics[width=0.7\textwidth]{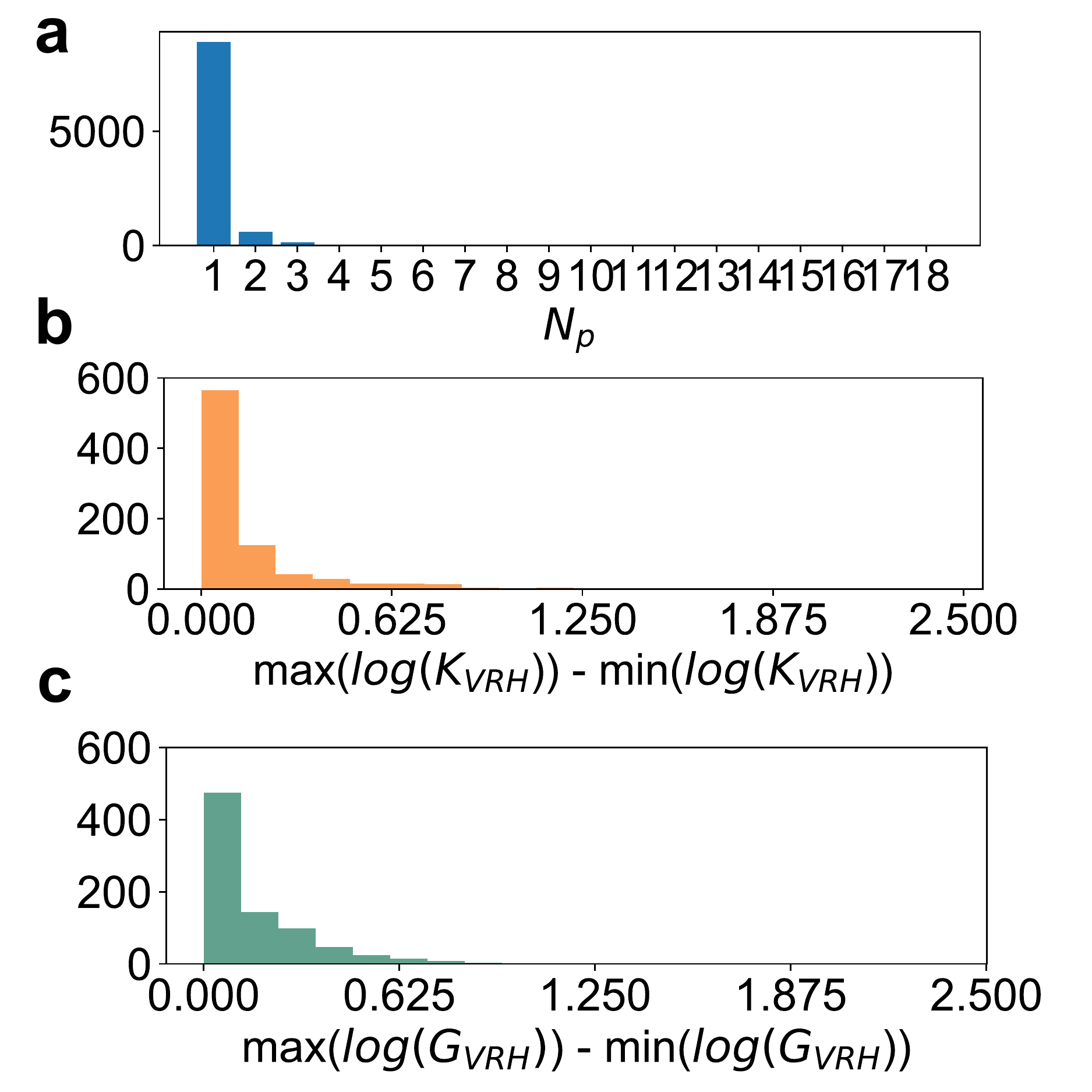}
\caption{Analysis of polymorph distribution in the elasticity dataset. (a) The distribution of the number of polymorphs $N_p$ for a given composition. The range distribution of $\log10(K)$ (b) and $\log10(G)$ (c) within polymorphs for compositions with $N_p > 1$. }
\label{fig:poly_counts}
\end{figure}

\begin{figure}
\centering
\includegraphics[width=0.95\textwidth]{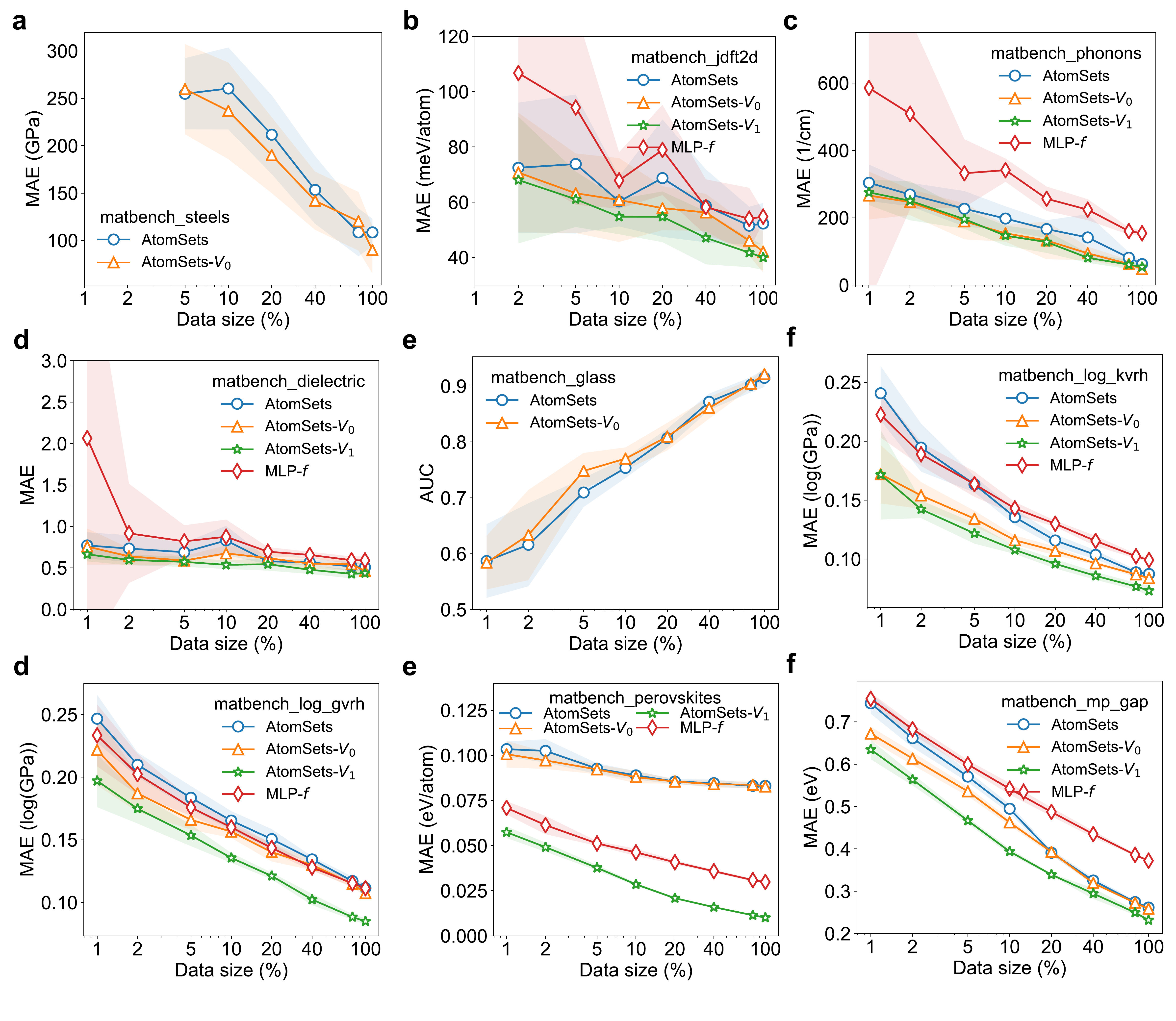}
\caption{Model convergence for AtomSets, AtomSets-$V_0$, AtomSets-$V_1$ and MLP-$f$ models for the nine datasets indicated by the legends.}
\label{fig:bench_conv}
\end{figure}

\begin{figure}
\centering
\includegraphics[width=0.95\textwidth]{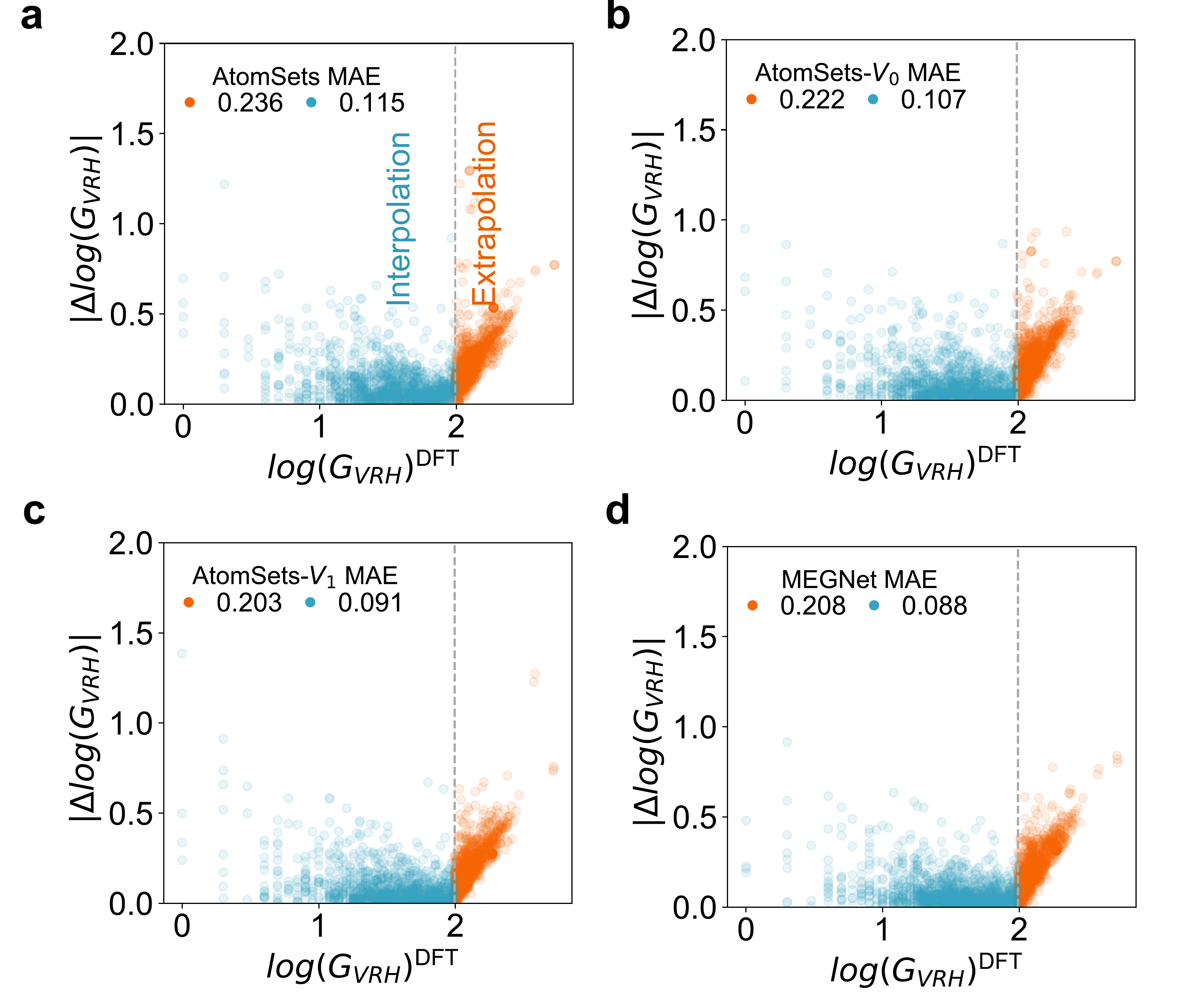}
\caption{Absolute differences in predicted and DFT $\log(G_{VRH})$, i.e., $|\Delta \log(G_{VRH})|$ against the DFT value range. The training and validation data are randomly sampled from the 0\% to 90\% (vertical dash line) target quantile range. Half of test data comes from the 90\%-100\% quantile (extrapolation) and the other half is from the same target range as the train-validation data (interpolation).}
\label{fig:logg}
\end{figure}

\clearpage
